\documentclass[twocolumn]{jpsj2}

\usepackage{times}

\usepackage{graphicx}

\title{Enhanced Kondo Effect in an Electron System
Dynamically Coupled with Local Optical Phonon}

\author{Takashi {\sc Hotta}}

\inst{Advanced Science Research Center,
Japan Atomic Energy Agency,
Tokai, Ibaraki 319-1195}

\recdate{\today}

\abst{
We discuss Kondo behavior of a conduction electron system
coupled with local optical phonon
by analyzing the Anderson-Holstein model
with the use of a numerical renormalization group (NRG) method.
There appear three typical regions due to the balance
between Coulomb interaction $U_{\rm ee}$ and
phonon-mediated attraction $U_{\rm ph}$.
For $U_{\rm ee}>U_{\rm ph}$, we observe the standard Kondo effect
concerning spin degree of freedom.
Since the Coulomb interaction is effectively reduced as
$U_{\rm ee}-U_{\rm ph}$,
the Kondo temperature $T_{\rm K}$ is increased
when $U_{\rm ph}$ is increased.
On the other hand, for $U_{\rm ee}<U_{\rm ph}$, there occurs
the Kondo effect concerning charge degree of freedom,
since vacant and double occupied states play roles of pseudo-spins.
Note that in this case, $T_{\rm K}$ is decreased
with the increase of $U_{\rm ph}$.
Namely, $T_{\rm K}$ should be maximized for $U_{\rm ee} \approx U_{\rm ph}$.
Then, we analyze in detail the Kondo behavior at $U_{\rm ee}=U_{\rm ph}$,
which is found to be explained by the polaron Anderson model
with reduced hybridization of polaron and
residual repulsive interaction among polarons.
By comparing the NRG results of the polaron Anderson model
with those of the original Anderson-Holstein model,
we clarify the Kondo behavior in the competing region
of $U_{\rm ee} \approx U_{\rm ph}$.
}

\kword{Kondo effect, Anderson-Holstein model, Polaron,
Numerical renormalization group method}

\begin{document}
\maketitle

%
%
\section{Introduction}

Kondo effect and its related phenomena have been currently investigated
intensively and extensively in the research field of condensed matter
physics,\cite{Kondo40} even after more than forty years have passed
since the pioneering work of Kondo in 1964.\cite{Kondo1}
It has been widely recognized that the Kondo-like phenomenon
generally occurs in a conduction electron system in which
a localized entity with internal degrees of freedom is embedded.
Then, a new mechanism of Kondo phenomenon with non-magnetic origin
has been potentially discussed, although the original Kondo effect
concerning local magnetic moment has been perfectly understood.

Concerning such non-magnetic Kondo effect,
Kondo himself has first considered a conduction electron system
which is coupled with a local double-well potential.\cite{Kondo2a,Kondo2b}
Two possibilities for electron position in the double-well potential
play roles of pseudo-spins and the Kondo-like behavior
is considered to appear in such a two-level system.
In fact, it has been shown that the two-level Kondo system
exhibits the same behavior as the magnetic Kondo effect.
\cite{Vladar1,Vladar2}
Recently, four- and six-level Kondo systems have been also analyzed,
\cite{Miyake,Hattori1,Hattori2}
in order to understand magnetically robust
heavy-fermion phenomenon observed in SmOs$_4$Sb$_{12}$.\cite{Sanada,Yuhasz}

The multi-level Kondo problem is considered to stem from
Kondo physics in electron-phonon systems.
For instance, the present author has discussed how Kondo-like
phenomenon occurs in a conduction electron system
coupled with local Jahn-Teller phonon.
\cite{Hotta1,Hotta2,Hotta3,Hotta4}
In order to overview the situation,
it is convenient to envisage the electron potential
in an adiabatic approximation,
although in actuality, the potential is not static,
but it dynamically changes to follow the electron motion.
When we simply ignore anharmonicity in the potential of
Jahn-Teller phonon, there exists continuous degeneracy
along the circle of the bottom of a Mexican-hat potential,
characteristic of Jahn-Teller system.
When we further take into account the effect of cubic anharmonicity,
three potential minima appear in the bottom of the Mexican-hat potential.
Then, we effectively obtain the three-level Kondo system.

It is quite meaningful to pursue a new possibility of
Kondo effect in the multi-level Kondo system
or in the Anderson model dynamically coupled
with local Jahn-Teller phonon.
On the other hand, it is also important to confirm
the fundamentals of Kondo physics in electron-phonon systems,
in parallel with the research of complex models
with close relation to actual materials.
Thus, we believe that it is useful to clarify the Kondo behavior
of a simple electron-phonon model.
In this sense, here we pick up the Anderson model coupled with
local optical phonon,
called the Anderson-Holstein model.\cite{Hewson,Jeon,Lee}
Concerning the origin of heavy-fermion phenomenon in SmOs$_4$Sb$_{12}$,
the periodic Anderson-Holstein model has been also analyzed,\cite{Ono}
and a mechanism of the mass enhancement due to electron-phonon
interaction has been addressed.

We note that the effect of Holstein phonon on the Kondo phenomenon
is considered to be limited, if we use the adiabatic approximation,
since the adiabatic potential would simply reduce the Coulomb interaction.
It is rather interesting to examine dynamical phonon effect
on the standard Kondo behavior concerning spin degree of freedom.
From a viewpoint of the relation with actual materials,
the Kondo behavior of the Anderson-Holstein model
in the anti-adiabatic region may have possible relevance with
the enhanced Kondo effect in molecular quantum dots.
\cite{qd1,qd2,qd3,qd4,qd5}
Since such a system is composed of light atoms,
relatively high frequency phonon may play an important
role for the Kondo physics
through the competition with Coulomb interaction.

In this paper, we analyze the Anderson-Holstein model
by using a numerical renormalization group (NRG) method.
The results are classified into three categories, depending on
Coulomb repulsion $U_{\rm ee}$ and phonon-mediated attraction
$U_{\rm ph}$.
For $U_{\rm ee}>U_{\rm ph}$, we easily understand that
the standard Kondo effect occurs,
since the repulsive interaction is still dominant,
even though it is effectively reduced as $U_{\rm ee}-U_{\rm ph}$.
Then, the Kondo temperature $T_{\rm K}$
is increased when $U_{\rm ph}$ is increased.
Note that in this paper, $T_{\rm K}$ is defined as a temperature
which exhibits a peak in the specific heat.
For $U_{\rm ee}<U_{\rm ph}$, we observe the Kondo effect
concerning charge degree of freedom,
since vacant and double occupied states play roles of pseudo-spins.
In this case, $T_{\rm K}$ is decreased with the increase of $U_{\rm ph}$.
Thus, $T_{\rm K}$ is considered to be maximized
for $U_{\rm ee} \approx U_{\rm ph}$.
We focus on the case of $U_{\rm ee}=U_{\rm ph}$ and 
the results are found to be understood by the polaron
Anderson model with reduced hybridization and
residual Coulomb interaction between polarons.
We discuss in detail the Kondo behavior of the Anderson-Holstein
model in the region of $U_{\rm ee}=U_{\rm ph}$ in comparison with
the polaron Anderson model.

The organization of this paper is as follows.
In Sec.~2, we introduce the Anderson-Holstein model
and provide a brief explanation of the NRG technique used here.
We also show the canonical transformation of the model
to diagonalize the phonon part,
in order to visualize the competition between Coulomb repulsion
$U_{\rm ee}$ and phonon-mediated attraction $U_{\rm ph}$.
In Sec.~3, we show our numerical results
for $U_{\rm ee} > U_{\rm ph}$ and $U_{\rm ee} < U_{\rm ph}$.
In order to analyze the Kondo temperature,
we introduce the effective $s$-$d$ models for both cases.
In Sec.~4, we discuss in detail the NRG results for
$U_{\rm ee} = U_{\rm ph}$.
For the purpose of intuitive understanding of the results,
we propose the polaron Anderson model.
Finally, in sec.~5, we briefly discuss the difference
in the Kondo effects between Holstein and Jahn-Teller phonons.
Throughout this paper, we use such units as $\hbar$=$k_{\rm B}$=1
and the energy unit is set as eV.

%
%
\section{Model and Method}

\subsection{Anderson-Holstein model}

Let us introduce the Anderson model coupled with local optical phonon.
The model Hamiltonian is expressed as
\begin{equation}
  H=\sum_{\mib{k}\sigma} \varepsilon_{\mib{k}}
    c_{\mib{k}\sigma}^{\dag} c_{\mib{k}\sigma}
   +\sum_{\mib{k}\sigma} (Vc_{\mib{k}\sigma}^{\dag}d_{\sigma}+{\rm h.c.})
   +H_{\rm loc},
\end{equation}
where $\varepsilon_{\mib{k}}$ is the dispersion of conduction electron,
$c_{\mib{k}\sigma}$ is an annihilation operator of conduction electron
with momentum $\mib{k}$ and spin $\sigma$,
$d_{\sigma}$ is an annihilation operator of localized electron
on an impurity site with spin $\sigma$, and
$V$ is the hybridization between conduction and localized electrons.
We choose $V$=0.25 and the energy unit is a half of the conduction
bandwidth, $D$, which is set as 1 eV throughout this paper.

The local term $H_{\rm loc}$ is given by
\begin{equation}
  H_{\rm loc}=U_{\rm ee}n_{\uparrow}n_{\downarrow}+\mu \rho+H_{\rm eph},
\end{equation}
where $U_{\rm ee}$ denotes Coulomb interaction,
$n_{\sigma}$=$d^{\dag}_{\sigma}d_{\sigma}$,
$\mu$ is a chemical potential,
and $\rho$=$n_{\uparrow}+n_{\downarrow}$.
We adjust $\mu$ appropriately to consider the half-filling case,
but the explicit value will be shown later.

The electron-phonon coupling term $H_{\rm eph}$ is given by
\begin{equation}
  H_{\rm eph}=g Q \rho + P^2/2+ \omega^2Q^2/2,
\end{equation}
where $g$ is the electron-phonon coupling constant,
$Q$ is normal coordinate of breathing mode,
and $P$ is the corresponding canonical momentum.
Note that the reduced mass of the breathing mode phonon
is set as unity.
Using the phonon operator $a$ defined through
$Q$=$(a+a^{\dag})/\sqrt{2\omega}$, we obtain
\begin{equation}
  H_{\rm eph} = 
  \omega\sqrt{\alpha}(a+a^{\dag}) \rho + \omega(a^{\dag}a+1/2),
\end{equation}
where $\alpha$ is the non-dimensional electron-phonon coupling
constant, given by $\alpha$=$g^2/(2\omega^3)$.
The phonon basis is given by
$|\ell \rangle$=$(a^{\dag})^{\ell}|0\rangle/\sqrt{\ell!}$,
where $\ell$ is the phonon number and
$|0\rangle$ is the vacuum state.
In actual calculations, the phonon basis is truncated
at a finite number, which is set as 400 in this paper.

\subsection{Numerical renormalization group method}

In this paper, the Anderson-Holstein model is analyzed by
a numerical renormalization group (NRG) method \cite{NRG},
in which momentum space is logarithmically discretized
to include efficiently the conduction electrons near the Fermi energy
and the conduction electron states
are characterized by ``shell'' labeled by $N$.
The shell of $N$=0 denotes an impurity site described by
the local Hamiltonian.
The Hamiltonian is transformed into the recursion form as
\begin{eqnarray}
  H_{N+1} = \sqrt{\Lambda}H_N + t_N \sum_\sigma
  (c_{N\sigma}^{\dag}c_{N+1\sigma}+c_{N+1\sigma}^{\dag}c_{N\sigma}),
\end{eqnarray}
where $\Lambda$ is a parameter for logarithmic discretization,
$c_{N\sigma}$ denotes the annihilation operator of conduction electron
in the $N$-shell, and $t_N$ indicates ``hopping'' of electron between
$N$- and $(N+1)$-shells, expressed by
\begin{eqnarray}
  t_N=\frac{(1+\Lambda^{-1})(1-\Lambda^{-N-1})}
  {2\sqrt{(1-\Lambda^{-2N-1})(1-\Lambda^{-2N-3})}}.
\end{eqnarray}
The initial term $H_0$ is given by
\begin{eqnarray}
  H_0=\Lambda^{-1/2}[H_{\rm loc}+\sum_{\sigma}
  V(c_{0\sigma}^{\dag}d_{\sigma}+d_{\sigma}^{\dag}c_{0\sigma})].
\end{eqnarray}

The free energy $F$ for local electron in each step is evaluated by
\begin{eqnarray}
   F = -T (\ln {\rm Tr} e^{-H_N/T} - \ln {\rm Tr} e^{-H_N^0/T}),
\end{eqnarray}
where a temperature $T$ is defined as $T$=$\Lambda^{-(N-1)/2}$
in the NRG calculation and $H_N^0$ denotes the Hamiltonian
without the hybridization term and $H_{\rm loc}$.
The entropy $S_{\rm imp}$ is obtained by
$S_{\rm imp}$=$-\partial F/\partial T$
and the specific heat $C_{\rm imp}$ is evaluated by
$C_{\rm imp}$=$-T\partial^2 F/\partial T^2$.
In the NRG calculation, we keep $m$ low-energy states
for each renormalization step.
In this paper, $\Lambda$ is set as 2.5 and we choose
$m$=5000, but for some case, it is necessary to increase $m$
up to 7500 to obtain convergent results.

In order to clarify the low-temperature properties,
we evaluate charge and charge susceptibilities,
given by, respectively,
\begin{eqnarray}
  \chi_{{\rm c}} =
  \frac{1}{Z} \sum_{n,m}
  \frac{e^{-E_n/T}-e^{-E_m/T}}{E_m-E_n}
  |\langle n | (\rho- \langle \rho \rangle) | m \rangle|^2,
\end{eqnarray}
and
\begin{eqnarray}
  \chi_{{\rm s}} =
  \frac{1}{Z} \sum_{n,m}
  \frac{e^{-E_n/T}-e^{-E_m/T}}{E_m-E_n}
  |\langle n | \sigma_{z} | m \rangle|^2,
\end{eqnarray}
where $E_n$ is the eigenenergy for the $n$-th eigenstate
$|n\rangle$ of $H$,
$Z$ is the partition function given by $Z$=$\sum_n e^{-E_n/T}$,
$\langle \rho \rangle$=
$(1/Z)\sum_{n} e^{-E_n/T} \langle n | \rho | n \rangle$,
and $\sigma_z$=$n_{\uparrow}-n_{\downarrow}$.
We perform the calculation in each step
by using the renormalized state.

\subsection{Lang-Firsov transformation}

The NRG calculations are performed for the Anderson-Holstein model,
but in order to grasp roughly the property of the model,
it is convenient to employ the Lang-Firsov canonical
transformation,\cite{LF}
defined through the change of an operator $A$ into
${\tilde A}$=$e^RAe^{-R}$ with
$R$=$\sqrt{\alpha}\rho(a^{\dag}-a)$.
Then, the Anderson-Holstein model is transformed into
\begin{equation}
 \label{HLF}
 \begin{split}
  {\tilde H} & =\sum_{\mib{k}\sigma} \varepsilon_{\mib{k}}
    c_{\mib{k}\sigma}^{\dag} c_{\mib{k}\sigma}
   +\sum_{\mib{k}\sigma} (Vc_{\mib{k}\sigma}^{\dag}d_{\sigma}X+{\rm h.c.}) \\
   &+(U_{\rm ee}-U_{\rm ph}) n_{\uparrow}n_{\downarrow}
   +(\mu-\alpha\omega)\rho+\omega(a^{\dag}a+1/2),
 \end{split}
\end{equation}
where $X$=$e^{\sqrt{\alpha}(a-a^{\dag})}$ and $U_{\rm ph}$=$2\alpha\omega$.
Note that the Coulomb repulsion is reduced by
the phonon-mediated attractive interaction.
We easily understand that for $U_{\rm ee}>U_{\rm ph}$,
the local state is doubly degenerate with spin degree of freedom,
while for $U_{\rm ee}<U_{\rm ph}$,
the vacant and double occupied states are degenerate at half-filling.
From the canonical transformation,
we obtain that the chemical potential at half-filling is given by
$\mu$=$-U_{\rm ee}/2+2\alpha\omega$.
In the following section, we will discuss the numerical results
in three regions as $U_{\rm ee}>U_{\rm ph}$, $U_{\rm ee}<U_{\rm ph}$,
and $U_{\rm ee}=U_{\rm ph}$.

\begin{figure}[t]
\centering
\includegraphics[width=7truecm]{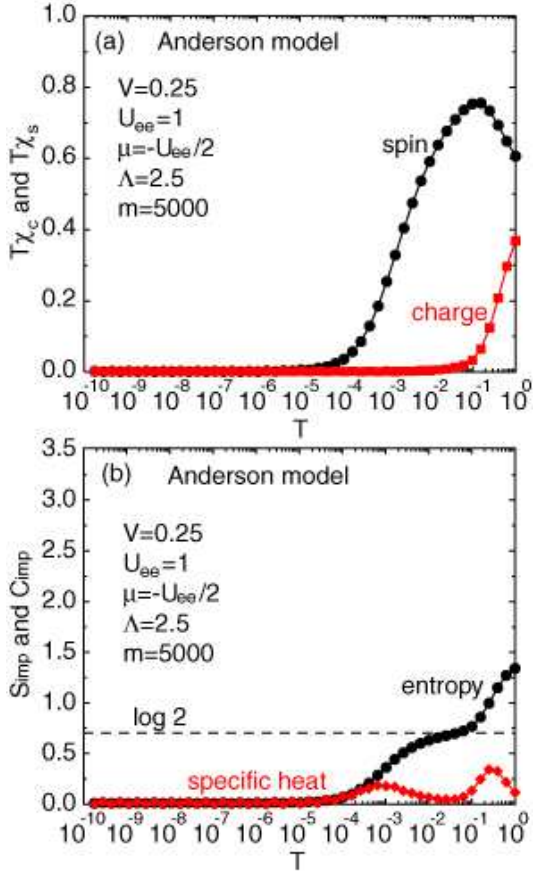}
\caption{(Color online)
(a) $T\chi_{\rm c}$ and $T\chi_{\rm s}$ vs. temperature
for the Anderson model with $U_{\rm ee}=1$.
(b) $S_{\rm imp}$ and $C_{\rm imp}$ vs. temperature
for the Anderson model with $U_{\rm ee}=1$.}
\end{figure}

%
%
\section{Kondo Behavior for $U_{\rm ee} \ne U_{\rm ph}$}

\subsection{Numerical results}

First let us briefly review the results of the Anderson model,
in order to clarify the effect of Holstein phonon.
In Fig.~1(a), we show $T\chi_{\rm c}$ and $T\chi_{\rm s}$
for $U_{\rm ee}$=1.
The charge susceptibility is rapidly suppressed due to
the effect of on-site Coulomb interaction, while the
spin susceptibility is increased.
As is well known, in the Kondo system,
the renormalization flow moves toward the strong-coupling regime
in which the spin susceptibility is suppressed,
through the local moment regime with enhanced spin susceptibility.

The Kondo behavior is clearly observed in the entropy and
the specific heat.
After the charge susceptibility is suppressed
around at a temperature in the order of $U_{\rm ee}$,
we can observe the local moment region with $\log 2$
between $0.01 < T < 0.1$.
Then, the entropy of $\log 2$ is gradually released and
around at $T \sim 10^{-4}$, it eventually goes to zero.
We can see a clear peak in the specific heat
around at $T \sim 10^{-3}$
due to the release of spin entropy $\log 2$.
We define the Kondo temperature $T_{\rm K}^{(0)}$
of the Anderson model as a lower peak in the specific heat.
It is well known that $T_{\rm K}^{(0)}$ is scaled by
$e^{-1/(2\rho_0 J_0)}$, where the exchange interaction $J_0$
is given by $J_0$=$4V^2/U_{\rm ee}$ and
$\rho_0$ is the density of states at the Fermi level.
Although a peak in the specific heat does not indicate
exactly the Kondo temperature,
the effect of the prefactor is simply ignored here,
since in this paper, we concentrate on the scaling relation
when the parameters of the model are changed.
Thus, we conventionally define $T_{\rm K}$ as the lower peak
in the specific heat throughout this paper.

\begin{figure}[t]
\centering
\includegraphics[width=7truecm]{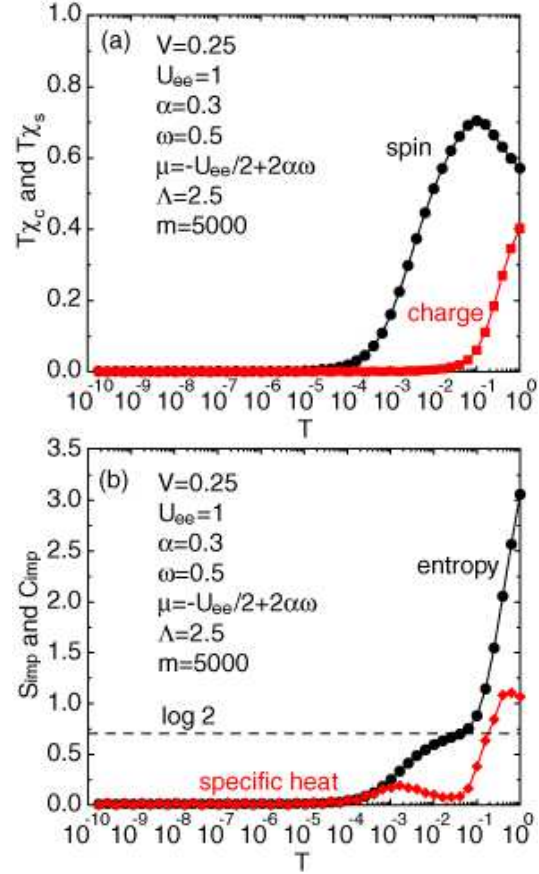}
\caption{(Color online)
(a) $T\chi_{\rm c}$ and $T\chi_{\rm s}$ vs. temperature
for $\alpha$=0.3, $\omega$=0.5, and $U_{\rm ee}$=1.
(b) $S_{\rm imp}$ and $C_{\rm imp}$
vs. temperature for the same parameters as (a).
}
\end{figure}

Now let us show the results of the Anderson-Holstein model.
First we consider the case in which the Coulomb interaction
is dominant.
In Fig.~2(a), we show the results for $T\chi_{\rm s}$ and
$T\chi_{\rm c}$ for $U_{\rm ee}$=1, $\alpha$=0.3, and $\omega$=0.5.
We note that in this case, $U_{\rm ph}$=0.3,
which is smaller than $U_{\rm ee}$.
We find that the charge susceptibility is rapidly
suppressed around at $T \sim 0.03$.
With decreasing temperature, the spin susceptibility is
suppressed around at $T \sim 10^{-4}$.
In Fig.~2(b), we show the entropy and specific heat
for the same parameters as in Fig.~2(a).
At high temperatures as $T>0.1$,
we observe the entropy larger than
$\log 4$ due to the low-lying phonon excitation states.
Passing through the narrow local moment region,
the entropy $\log 2$ is released and the peak appears in
the specific heat.
Note that the peak position is slightly shifted to
the higher-temperature side in comparison with Fig.~1(b).

The overall features of Figs.~2 are
quite similar to those of Figs.~1 except for
the high-temperature region,
since the situation is effectively understood by
the Anderson model with reduced Coulomb repulsion
$U_{\rm ee}-U_{\rm ph}$=0.7.
As we will discuss later, due to the decrease of
Coulomb interaction, the Kondo temperature is
increased in comparison with that of the Anderson model
with $U_{\rm ee}$=1.

\begin{figure}[t]
\centering
\includegraphics[width=7truecm]{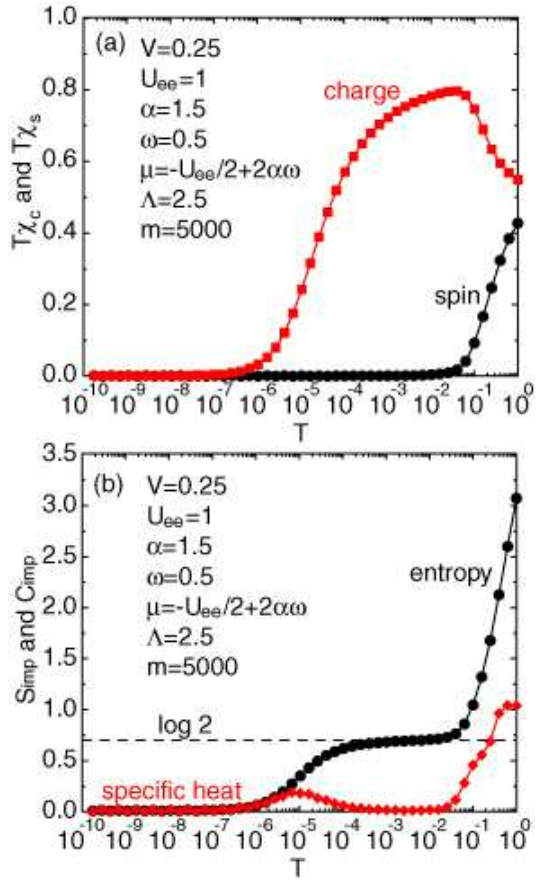}
\caption{(Color online)
(a) $T\chi_{\rm c}$ and $T\chi_{\rm s}$ vs. temperature
for $\alpha$=1.5, $\omega$=0.5, and $U_{\rm ee}$=1.
(b) $S_{\rm imp}$ and $C_{\rm imp}$ vs. temperature
for the same parameters as (a).
}
\end{figure}

Next we consider the situation with $U_{\rm ee}<U_{\rm ph}$.
In Fig.~3(a),
we show the results for $T\chi_{\rm s}$ and $T\chi_{\rm c}$
for $U_{\rm ee}$=1, $\alpha$=1.5, and $\omega$=0.5.
In this case, we obtain $U_{\rm ph}$=1.5,
leading to on-site attractive interaction with
$U_{\rm ee}-U_{\rm ph}$=$-0.5$.
Thus, in sharp contrast to Fig.~2(a), the spin susceptibility is
first rapidly suppressed around at $T \sim 0.1$.
With decreasing temperature, through the localized charge state,
the charge susceptibility vanishes around at $T \sim 10^{-6}$.

In Fig.~3(b), we depict the entropy and specific heat
for the same parameters as in Fig.~3(a).
At high temperatures, we again observe that
the entropy is larger than $\log 4$
due to the low-lying phonon states.
We find the localized charge region with $\log 2$
between $10^{-4} < T < 0.1$,
originating from the double degeneracy
of vacant and double occupied sites.
With decreasing temperature, the entropy $\log 2$ is
eventually released and a clear peak appears in
the specific heat.

Due to the effective attractive interaction,
we can find the Kondo behavior
concerning charge degree of freedom for $U_{\rm ee}<U_{\rm ph}$.
Since we are now considering the half-filling case,
as mentioned above, vacant and double occupied states
are degenerate and these states play roles of pseudo-spins.
Thus, we can understand the occurrence of
the Kondo-like behavior in the case of $U_{\rm ee}<U_{\rm ph}$.
The difference in the magnitude of the Kondo temperature
between the cases of $U_{\rm ee}>U_{\rm ph}$ and
$U_{\rm ee}<U_{\rm ph}$
will be discussed in the next subsection.

\subsection{Effective $s$-$d$ models}

In the previous subsection, we have shown the NRG results
for $U_{\rm ee}>U_{\rm ph}$ and $U_{\rm ee}<U_{\rm ph}$.
Here let us discuss analytic expressions for $T_{\rm K}$.
By using the second-order perturbation theory in terms of
the hybridization, we can derive the effective $s$-$d$ model
from the Anderson-Holstein model.
There appear the virtual second-order processes concerning
phonon excitations in addition to electronic excitations,
which affect on the exchange interactions.
The similar calculations have been done in the derivation of
the effective model from the Hubbard-Holstein model.
\cite{HottaHH1,HottaHH2,HottaHH3}

After some algebraic calculations, we obtain the effective $s$-$d$
models for $U_{\rm ee}>U_{\rm ph}$ and
$U_{\rm ee}<U_{\rm ph}$, respectively, as
\begin{equation}
 \begin{split}
  H^{\rm (s)}_{\rm s-d} &= \sum_{\mib{k}\sigma} \varepsilon_{\mib{k}}
  c_{\mib{k}\sigma}^{\dag} c_{\mib{k}\sigma} \\
  &+ \sum_{{\mib k},{\mib k'}}
  [J_+(c_{\mib{k}\uparrow}^{\dag} c_{\mib{k'}\uparrow}
  -c_{\mib{k}\downarrow}^{\dag} c_{\mib{k'}\downarrow})S_z \\
  &+J_+(c_{\mib{k}\downarrow}^{\dag} c_{\mib{k'}\uparrow}S_+
  +c_{\mib{k}\uparrow}^{\dag} c_{\mib{k'}\downarrow}S_-)],
 \end{split}
\end{equation}
and
\begin{equation}
 \begin{split}
  H^{\rm (c)}_{\rm s-d} &= \sum_{\mib{k}\sigma} \varepsilon_{\mib{k}}
  c_{\mib{k}\sigma}^{\dag} c_{\mib{k}\sigma} \\
  &+ \sum_{{\mib k},{\mib k'}}
  [J_+(c_{\mib{k}\uparrow}^{\dag} c_{\mib{k'}\uparrow}
  +c_{\mib{k}\downarrow}^{\dag} c_{\mib{k'}\downarrow}-1)\eta_z \\
  &+J_-(c_{\mib{k}\uparrow} c_{\mib{k'}\downarrow}\eta_+
  +c_{\mib{k}\downarrow}^{\dag} c_{\mib{k'}\uparrow}^{\dag}\eta_-)],
 \end{split}
\end{equation}
where $S_z$=$(n_\uparrow-n_\downarrow)/2$,
$S_{+}$=$d_{\uparrow}^{\dag}d_{\downarrow}$,
$S_{-}$=$d_{\downarrow}^{\dag}d_{\uparrow}$,
$\eta_z$=$(|2 \rangle \langle 2|-|0 \rangle \langle 0|)/2$,
$\eta_{+}$=$|2 \rangle \langle 0|$, and
$\eta_{-}$=$|0 \rangle \langle 2|$ with
$|2 \rangle$=$d^{\dag}_{\uparrow}d^{\dag}_{\downarrow}|0\rangle$.

\begin{figure}[t]
\centering
\includegraphics[width=7truecm]{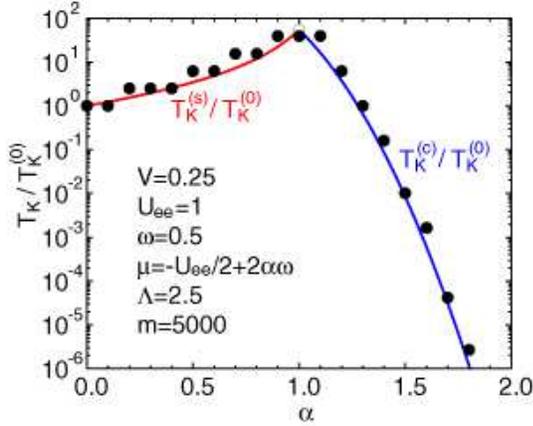}
\caption{(Color online)
Kondo temperature vs. $\alpha$ for $\omega$=0.5 and $U_{\rm ee}$=1.
Solid symbols denotes numerical results while solid curves
indicate analytic results of $T_{\rm K}^{\rm (c)}$ and
$T_{\rm K}^{\rm (s)}$.
The open circle at $\alpha$=1 indicates that
the analytic results cannot be used at this point.
}
\end{figure}

The exchange interactions are expressed as
\begin{equation}
 J_{\pm} = 4V^2 e^{-\alpha} \sum_{\ell=0}^{\infty}
  \frac{(\pm \alpha)^{\ell}}{\ell !}
  \frac{1}{ |U-U_{\rm ph}|+2\ell \omega}.
\end{equation}
We note that this expression does not hold around
at $U_{\rm ee}=U_{\rm ph}$.
Note also that the longitudinal and transverse parts of
$H^{\rm (c)}_{\rm s-d}$ are given by $J_+$ and $J_-$,
respectively, while for $H^{\rm (s)}_{\rm s-d}$,
both are given by $J_+$.
The difference between $J_+$ and $J_-$ clearly appears in
the asymptotic form for large $\alpha$ as
$J_+ \sim 1/\alpha$ and $J_- \sim e^{-2\alpha}/\alpha$.
Namely, in the strong electron-phonon coupling region,
$J_-$ decays very rapidly, while $J_+$ becomes small slowly.
The smallness of $J_-$ for large $\alpha$ originates from
the immobile nature of bi-polaron.
Single polaron can be relatively mobile in comparison
with bi-polaron.

Now we discuss the Kondo temperatures on the basis of
the effective $s$-$d$ models.
For the case of $U_{\rm ee}>U_{\rm ph}$,
$H_{\rm s-d}^{\rm (s)}$ is the isotropic $s$-$d$ model and
we can easily obtain
\begin{equation}
  \label{Tks}
  T^{\rm (s)}_{\rm K}=D {\rm exp}\Bigl(-\frac{1}{2\rho_0J_+}\Bigr).
\end{equation}
On the other hand, $H_{\rm s-d}^{\rm (c)}$ is the $s$-$d$ model
with anisotropic exchange interaction.
For this case, Shiba has obtained the explicit expression for
the binding energy ${\tilde E}$.\cite{Shiba}
When we define the Kondo temperature $T_{\rm K}$ as
$T_{\rm K}=-{\tilde E}$, we obtain
\begin{equation}
 \label{Tkc}
 T^{\rm (c)}_{\rm K}=D {\rm exp} \Biggl[
 \frac{-1}{2\rho_0\sqrt{J_+^2-J_-^2}}
 {\rm tanh}^{-1}\Bigl(\frac{\sqrt{J_+^2-J_-^2}}{J_+} \Bigr) \Biggr].
\end{equation}

In Fig.~4, we depict $T_{\rm K}/T_{\rm K}^{(0)}$ vs. $\alpha$.
Numerical results are shown by solid symbols.
Note that $T_{\rm K}$ as well as $T_{\rm K}^{(0)}$
in the numerical results is defined as a temperature
which shows the peak in the specific heat.
Analytic results for the $s$-$d$ models are depicted
by solid curves,
which indicate eqs.~(\ref{Tks}) and (\ref{Tkc}) divided by
$T_{\rm K}^{(0)}$.
We find that the numerical results agree well with
the analytic curves for $H_{\rm s-d}^{\rm (s)}$ and
$H_{\rm s-d}^{\rm (c)}$.
Note that the numerical results seem to scatter due to the effect
of discrete temperature $\Lambda^{-(N-1)/2}$, but
all the results are considered to be in the error-bars.

As mentioned above, for $U_{\rm ee}<U_{\rm ph}$,
the effective $s$-$d$ model becomes highly anisotropic
for large $\alpha$.
Since the transverse part is exponentially small,
the interaction part of $H_{\rm s-d}^{\rm (c)}$ becomes Ising-like
and the Kondo temperature is rapidly suppressed.
Thus, the plot of $T_{\rm K}$ vs. $\alpha$ is asymmetric
at the center of $\alpha$=1 ($U_{\rm ee}$=$U_{\rm ph}$).
Note also that two analytic curves seem to converge to
a value similar to the numerical result at $\alpha$=1,
but as mentioned above, the expression of $T_{\rm K}$ 
does not hold around at $U_{\rm ee}$=$U_{\rm ph}$.
The Kondo behavior in the repulsion-attraction competing
region with $U_{\rm ee} \approx U_{\rm ph}$ will
be separately discussed in the next section.

%
%

\begin{figure}[t]
\centering
\includegraphics[width=7truecm]{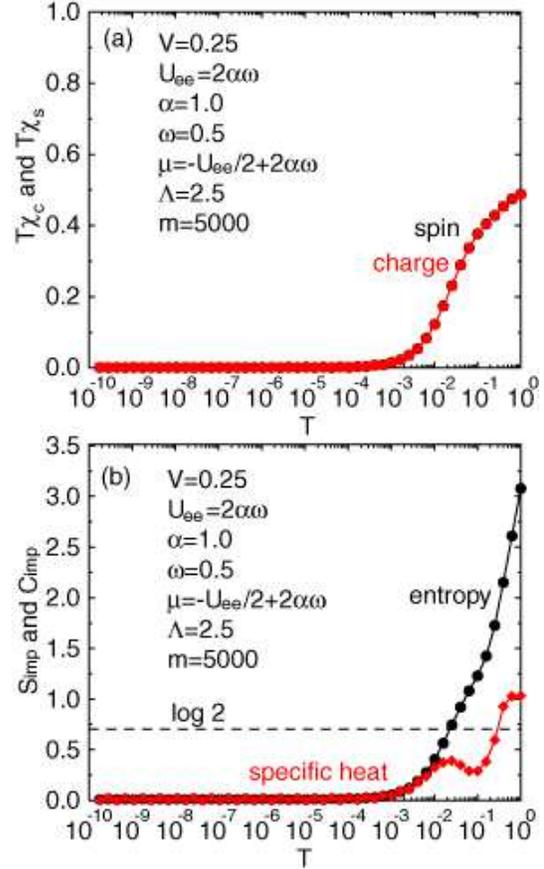}
\caption{(Color online)
(a) $T\chi_{\rm c}$ and $T\chi_{\rm s}$ vs. temperature
for $\alpha$=1 and $\omega$=0.5 with $U_{\rm ee}$=$2\alpha\omega$.
(b) $S_{\rm imp}$ and $C_{\rm imp}$
vs. temperature for the same parameters as (a).
}
\end{figure}

\section{Kondo Behavior in the Competing Region}

Now we focus on the case of $U_{\rm ee}=U_{\rm ph}$.
In Fig.~5(a), we show the results for $T\chi_{\rm s}$ and $T\chi_{\rm c}$
for $\alpha$=1, $U_{\rm ee}$=1, and $\omega$=0.5.
As naively expected from the cancellation of the on-site interaction,
the situation is considered to be understood by
the non-interacting model.
In fact, as shown in Fig.~5(a), it is very difficult to distinguish
the charge and spin susceptibilities,
indicating that the interaction effect is virtually ignored.

In Fig.~5(b), the entropy and specific heat are shown
for the same parameters as in Fig.~5(a).
At high temperatures as $T > 0.1$, in common with the cases of
$U_{\rm ee} \ne U_{\rm ph}$, we find that the entropy is larger
than $\log 4$ due to the low-lying phonon states.
We note that the entropy gradually goes to zero without showing
the region of $\log 2$.
We can see the peak in the specific heat around at $T \sim 0.01$,
which has been assigned as the Kondo temperature in Fig.~4.
However, this Kondo-like behavior is not due to Coulomb interaction,
but it originates from the hybridization
in the non-interacting Anderson model.
This point will be discussed later again.

\begin{figure}[t]
\centering
\includegraphics[width=7truecm]{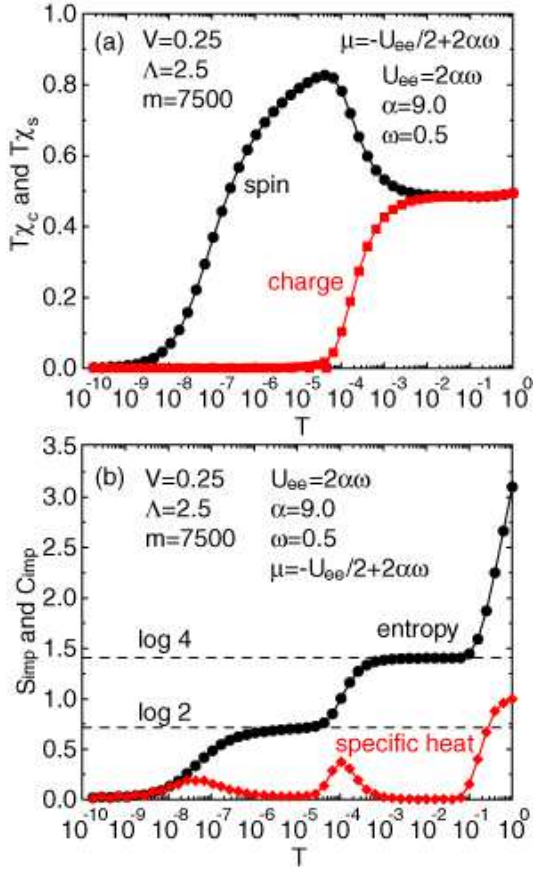}
\caption{(Color online)
(a) $T\chi_{\rm c}$ and $T\chi_{\rm s}$ vs. temperature
for $\alpha$=9 and $\omega$=0.5 with $U_{\rm ee}$=$2\alpha\omega$.
(b) $S_{\rm imp}$ and $C_{\rm imp}$
vs. temperature for the same parameters as (a).
Note that in this case, we increase $m$ up to 7500,
since convergence becomes worse due to the strong-coupling nature.
}
\end{figure}

Next we increase the value of $\alpha$ by keeping the
relation of $U_{\rm ee}=U_{\rm ph}$.
In Fig.~6(a),
we show the results for $T\chi_{\rm s}$ and $T\chi_{\rm c}$
for $\alpha$=9, $U_{\rm ee}$=$2\alpha\omega$, and $\omega$=0.5.
From the viewpoint of the comparison with actual materials,
the value of $\alpha$=9 seems to be unrealistically large,
but we consider such a situation in order to complete the
discussion from the theoretical viewpoint.
As naively expected from the cancellation of on-site interaction,
the situation can be understood from the non-interacting model,
but in the strong-coupling case, the results clearly show
the interaction effect.
For $T>10^{-3}$, $T\chi_{\rm c}$ and $T\chi_{\rm s}$ takes
a constant value of 0.5.
This region can be interpreted as the free orbital regime,
described by the conduction electron systems.\cite{NRG}
Around at $T \sim 10^{-3}$, spin and charge response
begin to be separated from each other.
The charge susceptibility is rapidly suppressed, while
the spin susceptibility grows up,
suggesting the local moment regime.
After that, the spin susceptibility is gradually suppressed
and it eventually goes to zero around at $T \sim 10^{-9}$,
entering the strong-coupling regime.

The above behavior can be clearly found in the entropy
and the specific heat, as shown in Fig.~6(b).
The high-temperature behavior is almost the same as
that in Fig.~5(b).
For $0.1<T<10^{-4}$, we clearly observe the region of $\log 4$
due to the four-fold degeneracy
of vacant, spin-up, spin-down, and double occupied states,
corresponding to the free-orbital regime.
Then, the entropy of $\log 2$ relevant to charge degree of freedom
is released and a peak is formed in the specific heat.
Finally, the residual entropy of $\log 2$ for spin degree of
freedom is released and we find another peak in the specific heat.

From the above numerical results,
we deduce that the case of $U_{\rm ee}=U_{\rm ph}$ is
described by the Anderson model with
effective on-site repulsive interaction.
From ${\tilde H}$, eq.~(\ref{HLF}), 
we propose the polaronic Anderson model, given by
\begin{equation}
  \begin{split}
  H_{\rm eff} &= \sum_{{\bf k}\sigma} \varepsilon_{\bf{k}}
      c_{\bf{k}\sigma}^{\dag} c_{\bf{k}\sigma}
  +\sum_{\bf{k}\sigma}(V_{\rm eff}
      c_{\bf{k}\sigma}^{\dag}d_{\sigma}+{\rm h.c.}) \\
  &+U_{\rm eff} n_{\uparrow} n_{\downarrow}
   +\mu_{\rm eff} \rho,
  \end{split}
\end{equation}
where $V_{\rm eff}$ is the hybridization for polarons,
$U_{\rm eff}$ is the residual Coulomb interaction
among polarons,
and $\mu_{\rm eff}$ is the chemical potential for polarons,
given by $\mu_{\rm eff}$=$-U_{\rm eff}/2$.

By taking the average over the zero-phonon state,
we obtain $V_{\rm eff}$ as
$V_{\rm eff}$=$V\langle 0 | X |0 \rangle$=$V e^{-\alpha/2}$.
However, we cannot hit upon an idea to derive the analytic
form of $U_{\rm eff}$.
Then, we resort to a numerical method to determine $U_{\rm eff}$
so as to reproduce the average value of double occupancy
$\langle n_{\uparrow} n_{\downarrow} \rangle$ of $H$.

\begin{figure}[t]
\centering
\includegraphics[width=7truecm]{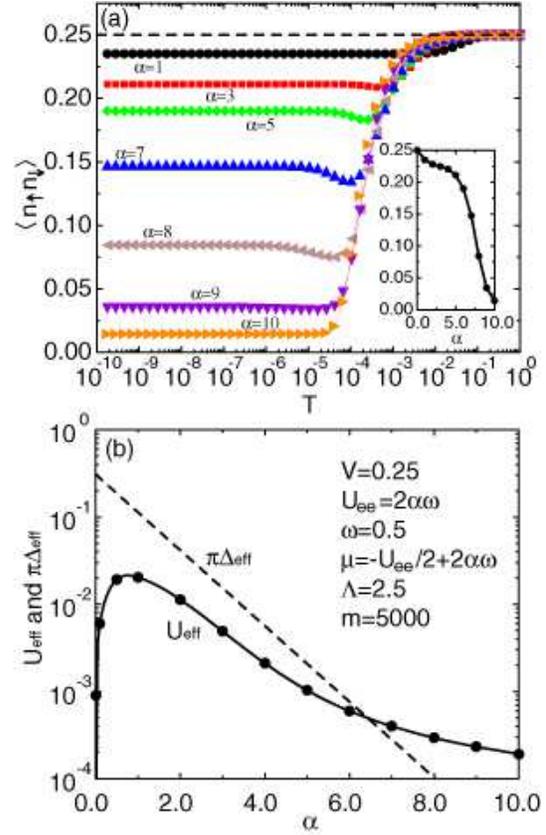}
\caption{(Color online)
(a) $\langle n_{\uparrow} n_{\downarrow} \rangle$ vs.
temperature for several values of $\alpha$
with $\omega$=0.5 and $U_{\rm ee}$=$2\alpha\omega$.
Inset shows $\langle n_{\uparrow} n_{\downarrow} \rangle$
at the lowest temperature in the present NRG calculations.
(b) $\Delta_{\rm eff}$ and $U_{\rm eff}$ vs. $\alpha$
for $\omega$=0.5 and $U_{\rm ee}$=$2\alpha\omega$.
}
\end{figure}

In Fig.~7(a), we show $\langle n_{\uparrow} n_{\downarrow} \rangle$
for several values of $\alpha$ with $U_{\rm ee}$=$2\alpha\omega$.
In the inset, we depict $\langle n_{\uparrow} n_{\downarrow} \rangle$
at the lowest temperature which we can reach in the present
NRG calculation.
Irrespective of the vales of $\alpha$,
in the high-temperature region as $T>10^{-3}$,
$\langle n_{\uparrow} n_{\downarrow} \rangle$ takes the value
near the non-interacting one, 0.25.
When we further decrease the temperature,
for $\alpha < 6$, $\langle n_{\uparrow} n_{\downarrow} \rangle$
is not largely suppressed and it keeps the value about 0.2
even at low temperatures.
However, for $\alpha > 6$, it is rapidly suppressed
and takes the value much smaller than 0.25.
In fact, as shown in the inset, the low-temperature value of
$\langle n_{\uparrow} n_{\downarrow} \rangle$
is rapidly suppressed around at $\alpha \sim 6$.

Readers may feel it strange that the double occupancy is
so suppressed, in spite of the fact that the on-site
interaction exactly vanishes at $U_{\rm ee}=U_{\rm ph}$.
As mentioned in the explanation of the effective $s$-$d$ model,
bi-polaron has immobile nature in comparison with
single polaron for large $\alpha$.
Due to the difference in the mobility between polaron and bi-polaron,
single polaron state has the energy gain of the
exchange interaction, while bi-polaron cannot.
Thus, the single polaron state is favored in the
strong-coupling region, leading to the effective occurrence
of the residual repulsion between polarons.

By changing the values of $U_{\rm eff}$ for each values of $\alpha$,
we repeat the NRG calculations for $H_{\rm eff}$,
until we can reproduce the low-temperature value of
$\langle n_{\uparrow} n_{\downarrow} \rangle$ of $H$.
The results are shown by solid symbols in Fig.~7(b).
In order to understand intuitively the extent of correlation effect,
we also depict the line of $\pi \Delta_{\rm eff}$,
where $\Delta_{\rm eff}$ is the width of the virtual bound state,
expressed as
$\Delta_{\rm eff}$=$\pi\rho_0 V_{\rm eff}^2$=$\pi\rho_0V^2e^{-\alpha}$.
Note that the expansion parameter of the Anderson model
is given by $U_{\rm eff}/(\pi\Delta_{\rm eff})$.
\cite{Yamada1,Yamada2,Yamada3,Yamada4}
For $\alpha<6.5$, we observe that $U_{\rm eff}$ is smaller than
$\pi \Delta_{\rm eff}$, suggesting that
the peak in the specific heat should be determined by
$\Delta_{\rm eff}$.
For large $\alpha$, on the other hand, $U_{\rm eff}$ is still small,
but $\Delta_{\rm eff}$ is exponentially reduced,
as observed in Fig.~7(b).
Then, in this region, we expect that the correlation effect
becomes significant.

\begin{figure}[t]
\centering
\includegraphics[width=7truecm]{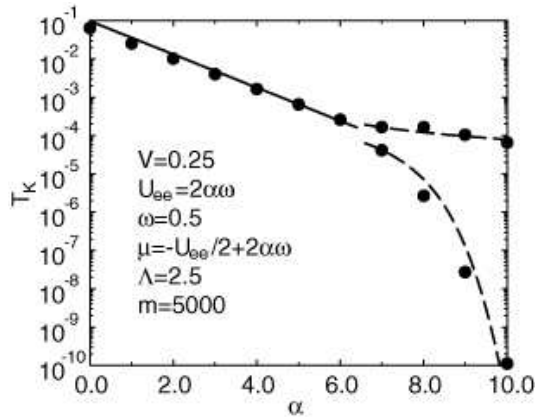}
\caption{$T_{\rm K}$ vs. $\alpha$ for
$\omega$=0.5 and $U_{\rm ee}$=$2\alpha\omega$.
}
\end{figure}

In Fig.~8, we summarize the numerical results
for the peak temperature in the specific heat
with the fitting curve deduced from
the polaron Anderson model.\cite{HottaSCES}
For $\alpha<6.5$, as described above,
the NRG results are simply scaled by $\Delta_{\rm eff}$.
For $\alpha>6.5$, as typically found in Figs.~6,
we obtain clear two peaks in the specific heat.
The higher peak concerning charge degree of freedom should
be characterized by $U_{\rm eff}$,
while the lower one indicates the characteristic
temperature of the standard Kondo effect, given by
$e^{-1/(2\rho_0 J_{\rm eff})}$
with $J_{\rm eff}$=$4 V_{\rm eff}^2 / U_{\rm eff}$.
By adjusting appropriately numerical prefactors
at $\alpha$=7, we can actually fit the NRG results by
$U_{\rm eff}$ and $e^{-1/(2\rho_0 J_{\rm eff})}$.
We believe that the success of this fitting suggests
the effectiveness of the polaron Anderson model,
which describes the low-energy states of
the Anderson-Holstein model
in the competing region of $U_{\rm ee} \approx U_{\rm ph}$.

%
%
\section{Discussion and Summary}

We have clarified the Kondo behavior of the Anderson-Holstein
model by using the NRG method.
It has been clearly shown that the results are categorized into
three classes labeled by $U_{\rm ee}>U_{\rm ph}$, $U_{\rm ee}<U_{\rm ph}$,
and $U_{\rm ee} \approx U_{\rm ph}$.
For $U_{\rm ee}>U_{\rm ph}$, the standard magnetic Kondo phenomenon
occurs with the reduced Coulomb interaction,
suggesting the increase of the Kondo temperature
in comparison with that of the Anderson model without
the electron-phonon interaction.
For $U_{\rm ee}<U_{\rm ph}$, the charge Kondo effect occurs
and the Kondo temperature is decreased with the increase of
electron-phonon coupling constant, since the relevant exchange
interaction is decreased when we increase $\alpha$.
Note that the effective $s$-$d$ model for this case becomes highly
anisotropic, since the transverse part is related to the bi-polaron
motion with immobile nature in comparison with single polaron.

Around at $U_{\rm ee} \approx U_{\rm ph}$,
the Kondo temperature is maximized, but for small $\alpha$,
the characteristic energy is simply given by the width of
the virtual bound state of the non-interacting Anderson model.
In this sense, we should not call the peak in the specific
heat as the Kondo temperature for small $\alpha$, but
the Kondo-like behavior occurs at relatively high temperature,
when the strong electron-phonon
interaction competes with Coulomb interaction.
We may consider possible relevance of this scenario
to the enhanced Kondo temperature observed in molecular quantum dots.
For larger $\alpha$, the effect of residual polaron repulsion
becomes significant, since the polaron hybridization is exponentially
suppressed.
Then, we have found the Kondo singlet formation,
through the free-orbital and local moment regimes.

Finally, we provide a comment on the different effect of
the phonon mode on the Kondo temperature.
In this paper, we have concentrated on the Holstein phonon and
stressed the enhancement of the Kondo temperature.
However, this enhancement depends on the feature of
relevant phonon.
For instance, for the case of Jahn-Teller phonon,
the Kondo temperature is monotonically decreased
with the increase of the electron-phonon coupling constant.
\cite{Hotta1,Hotta2}
For Jahn-Teller phonon, the local electron-phonon state has
double degeneracy originating from clockwise and anti-clockwise
rotational phonon mode.
Such geometrical degree of freedom is characterized by $J$=$\pm 1/2$,
where $J$ is total angular moment, composed of electron orbital
and phonon angular moments.
The conduction electrons must screen phonon angular moment
in addition to electron orbital moment
to form the singlet ground state with $J$=0.
Thus, the Kondo temperature is decreased when we increase $\alpha$.

In summary, we have discussed the Kondo effect of
the Anderson-Holstein model.
We have observed that the Kondo behavior can be explained
by the isotropic $s$-$d$ model for $U_{\rm ee}>U_{\rm ph}$,
the polaron Anderson model for $U_{\rm ee} \approx U_{\rm ph}$,
and the anisotropic $s$-$d$ model for $U_{\rm ee}<U_{\rm ph}$.
The Kondo behavior has been found to be enhanced
when $U$ competes with $U_{\rm ph}$.

\section*{Acknowledgement}

The author thanks K. Kubo, H. Onishi, K. Ueda, and Y. Kuramoto
for discussions and comments.
This work has been supported by a Grant-in-Aid for Scientific Research
in Priority Area ``Skutterudites'' under the contract No.~18027016
from the Ministry of Education, Culture, Sports, Science, and
Technology of Japan.
The author has been also supported by a Grant-in-Aid for
Scientific Research (C) under the contract No.~18540361
from Japan Society for the Promotion of Science.
The computation in this work has been done using the facilities
of the Supercomputer Center of Institute for Solid State Physics,
University of Tokyo.



\begin{thebibliography}{99}

\bibitem{Kondo40}
Kondo effect and its related phenomena have been reviewed
in J. Phys. Soc. Jpn. {\bf 74} (2005) 1-238.

\bibitem{Kondo1}
J. Kondo: Prog. Theor. Phys. {\bf 32} (1964) 37.

\bibitem{Kondo2a}
J. Kondo: Physica {\bf 84}B (1976) 40.

\bibitem{Kondo2b}
J. Kondo: Physica {\bf 84}B (1976) 207.

\bibitem{Vladar1}
K. Vladar and A. Zawadowski:
Phys. Rev. B {\bf 28} (1983) 1564.

\bibitem{Vladar2}
K. Vladar and A. Zawadowski:
Phys. Rev. B {\bf 28} (1983) 1582.

\bibitem{Miyake}
S. Yotsuhashi, M. Kojima, H. Kusunose and K. Miyake:
J. Phys. Soc. Jpn. {\bf 74} (2005) 49.

\bibitem{Hattori1}
K. Hattori, Y. Hirayama and K. Miyake:
J. Phys. Soc. Jpn. {\bf 74} (2005) 3306.

\bibitem{Hattori2}
K. Hattori, Y. Hirayama and K. Miyake:
{\it Proc. 5th Int. Symp. ASR-WYP-2005: Advances in the Physics and
Chemistry of Actinide Compounds},
J. Phys. Soc. Jpn. {\bf 75} (2006) Suppl., p. 238.

\bibitem{Sanada}
S. Sanada, Y. Aoki, H. Aoki, A. Tsuchiya, D. Kikuchi, H. Sugawara
and H. Sato: J. Phys. Soc. Jpn. {\bf 74} (2005) 246.

\bibitem{Yuhasz}
W. M. Yuhasz, N. A. Frederick, P.-C. Ho, N. P. Butch, B. J. Taylor,
T. A. Sayles, M. B. Maple, J. B. Betts, A. H. Lacerda, P. Rogl
and G. Giester: Phys. Rev. B {\bf 71} (2005) 104402.

\bibitem{Hotta1}
T. Hotta: Phys. Rev. Lett. {\bf 96} (2006) 197201.

\bibitem{Hotta2}
T. Hotta: J. Phys. Soc. Jpn. {\bf 76} (2007) 023705.

\bibitem{Hotta3}
T. Hotta: J. Magn. Magn. Mater. {\bf 310} (2007) 1691.

\bibitem{Hotta4}
T. Hotta: J. Phys. Soc. Jpn. {\bf 76} (2007) 034713.

\bibitem{Hewson}
A. C. Hewson and D. Meyer:
J. Phys.: Condens. Matter. {\bf 14} (2002) 427.

\bibitem{Jeon}
G. S. Jeon, T. -H. Park and H. -Y. Choi:
Phys. Rev. B {\bf 68} (2003) 045106.

\bibitem{Lee}
H. C. Lee and H. -Y. Choi:
Phys. Rev. B {\bf 69} (2004) 075109.

\bibitem{Ono}
K. Mitsumoto and Y. \=Ono:
Physica C {\bf 426-431} (2005) 330.

\bibitem{qd1}
L. H. Yu, Z. K. Keane, J. W. Ciszek, L. Cheng, M. P. Stewart,
J. M. Tour and D. Natelson:
Phys. Rev. Lett. {\bf 93} (2004) 266802.

\bibitem{qd2}
L. H. Yu, Z. K. Keane, J. W. Ciszek, L. Cheng, J. M. Tour, T. Baruah,
M. R. Pederson and D. Natelson:
Phys. Rev. Lett. {\bf 95} (2005) 245803.

\bibitem{qd3}
P. S. Cornaglia, H. Ness and D. R. Grempel:
Phys. Rev. Lett. {\bf 93} (2004) 147201.

\bibitem{qd4}
P. S. Cornaglia, D. R. Grempel and H. Ness:
Phys. Rev. B {\bf 71} (2005) 075320.

\bibitem{qd5}
P. S. Cornaglia and D. R. Grempel:
Phys. Rev. B {\bf 71} (2005) 245326.

\bibitem{NRG}
H. R. Krishna-murthy, J. W. Wilkins and K. G. Wilson:
Phys. Rev. B {\bf 21} (1980) 1003.

\bibitem{LF}
I. G. Lang and Yu. A . Firsov:
Zh. Eksp. Theor. Fiz. {\bf 43} (1962) 1843
[Sov. Phys. -JETP {\bf 16} (1963) 1301].

\bibitem{HottaHH1}
T. Hotta and Y. Takada:
Phys. Rev. Lett. {\bf 76} (1996) 3180. 

\bibitem{HottaHH2}
T. Hotta and Y. Takada:
J. Phys. Soc. Jpn. {\bf 65} (1996) 2922.

\bibitem{HottaHH3}
T. Hotta and Y. Takada:
Phys. Rev. B {\bf 56} (1997) 13916.

\bibitem{Shiba}
H. Shiba: Prog. Theor. Phys. {\bf 43} (1970) 601.

\bibitem{Yamada1}
K. Yamada and K. Yoshida: Prog. Thoer. Phys. Suppl. {\bf 46} (1970) 244.

\bibitem{Yamada2}
K. Yamada: Prog. Thoer. Phys. {\bf 53} (1975) 970.

\bibitem{Yamada3}
K. Yoshida and K. Yamada: Prog. Thoer. Phys. {\bf 53} (1975) 1286.

\bibitem{Yamada4}
K. Yamada: Prog. Thoer. Phys. {\bf 54} (1975) 316.

\bibitem{HottaSCES}
T. Hotta: submitted to Physica B.

\end{thebibliography}
\end{document}